\documentclass[aip, amsmath,amssymb, reprint]{revtex4-1}
\usepackage{amsmath}
\usepackage{graphicx}
\usepackage{color}
\usepackage[colorlinks=true,bookmarks=false,citecolor=blue,urlcolor=blue]{hyperref}
\newcommand{\be}{\begin{equation}}
\newcommand{\ee}{\end{equation}}

\newcommand{\revision}{}

\newcommand{\pcsadd}{Center for Theoretical Physics of Complex Systems, Institute for Basic Science, Daejeon 34126, Korea}
\newcommand{\ustadd}{Basic Science Program, Korea University of Science and Technology, Daejeon 34113, Korea}
\newcommand{\sjtuadd}{State Key Laboratory of Advanced Optical Communication Systems and Networks,\\ School of Physics and Astronomy, Shanghai Jiao Tong University, Shanghai, 200240, China}

\begin{document}

\title{Topological phases in ring resonators: recent progress and future prospects}

\author{Daniel Leykam}
\altaffiliation{Present address: Centre for Quantum Technologies, National University of Singapore, 3 Science Drive 2, Singapore 117543}
\affiliation{\pcsadd}
\affiliation{\ustadd}
\email{daniel.leykam@gmail.com}

\author{Luqi Yuan}
\affiliation{\sjtuadd}

\email{yuanluqi@sjtu.edu.cn}

\date{\today}

\begin{abstract}
Topological photonics has emerged as a novel paradigm for the design of electromagnetic systems from microwaves to nanophotonics. Studies to date have largely focused on the demonstration of fundamental concepts, such as non-reciprocity and waveguiding protected against fabrication disorder. Moving forward, there is a pressing need to identify applications where topological designs can lead to useful improvements in device performance. Here we review applications of topological photonics to ring resonator-based systems, including one- and two-dimensional resonator arrays, and dynamically-modulated resonators. We evaluate potential applications such as quantum light generation, disorder-robust delay lines, and optical isolation, as well as future research directions and open problems that need to be addressed.
\end{abstract}


\maketitle

\section{Introduction}

Demand for miniaturised optical components such as waveguides and lenses that can be incorporated into compact photonic devices is pushing fabrication techniques to their limits. Continued progress will require new approaches to minimise the detrimental influence of fabrication imperfections and disorder. Topological photonics is a young sub-field of photonics which seeks to address this challenge using novel design approaches inspired by exotic electronic condensed matter materials such as topological insulators~\cite{Ozawa2019,Khanikaev2017}. Loosely speaking, topological systems provide a systematic way to create disorder-robust modes or observables using a collection of imperfect components or modes. For example, certain classes of ``topologically nontrivial'' systems exhibit special edge modes which can propagate reliably without backscattering even in the presence of strongly scattering defects, forming the basis for superior optical waveguides.

One natural setting where this robustness can potentially be useful is in the design of integrated photonic circuits~\cite{You2017}, where the strong light confinement brings sensitivity to nanometer-scale fabrication imperfections. However, there is not yet any disruptive killer application where topological photonic devices have achieved superior performance compared to mature design paradigms, despite growing interest in topological photonics since seminal works published in 2008~\cite{Haldane2008,Wang2008}. To help bridge this gap, several reviews have been published recently, some providing comprehensive surveys of topological photonic systems~\cite{Ozawa2019,Khanikaev2017,Yuan2018,Ozawa2019NRP}, and others focusing on specific applications such as incorporating topological concepts into active devices such as lasers~\cite{active_review}, nanophotonics~\cite{Rider2019}, non-reciprocal devices~\cite{Chen2018review}, and nonlinear optical processes~\cite{nonlinear_review}.

The aim of this brief review is to complement these recent surveys with a concise introduction to applications of ideas from topological photonics to optical ring resonator-based systems. Ring resonators are a versatile and important ingredient of integrated photonic circuits, as they can be used as compact filters, sensors, and delay lines, and as a means of enhancing nonlinear optical effects~\cite{Bogaerts2012}. However, active tuning is typically required to compensate for resonance shifts induced by various perturbations, increasing device complexity and energy consumption~\cite{Zhou2015,Thomson2016}. We will discuss some of the ways in which topological designs may lead to superior devices with improved reliability. We will focus on topological systems formed by coupling together multiple resonators to form a lattice, or by considering coupling between multiple resonances of a single ring using external modulation or nonlinearity. In both cases the lattice formed by the coupled modes can be designed to have topological band structures and achieve protection against certain kinds of disorder. 

The outline of this article is as follows: Sec.~\ref{sec:background} starts with a brief overview of the basic concepts underlying topological photonics and ring resonators. Next, we review the implementation of topological photonics using arrays of coupled ring resonators in Sec.~\ref{sec:lattices}. Sec.~\ref{sec:modulation} discusses how we can use dynamic modulation as a novel degree of freedom for the engineering of topological effects in ring resonators. We discuss future research directions and promising potential applications of topological ring resonators in Sec.~\ref{sec:future}, before concluding with Sec.~\ref{sec:conclusion}.

\section{Background}
\label{sec:background}

\subsection{Topological photonics}
\label{sec:topology}

The key idea underlying topological photonics is the bulk-boundary correspondence, which states that the topological properties of the photonic band structure of a bulk periodic medium can be related to the appearance of robust modes localized to edges or domain walls of the system~\cite{Ozawa2019,Khanikaev2017}. These ``topologically-protected'' edge modes have very different properties compared to conventional defect modes. For example, as they arise due to the topological properties of the bulk photonic band structure, they are robust against certain classes of local perturbations at the edge or domain wall and can only be removed by large perturbations capable of closing the bulk band gap. Thus, topology allows us to create special robust boundary modes protected by a higher-dimensional bulk. 

For the purposes of this review, two classes of topological modes are of interest. One-dimensional media can exhibit topologically-protected modes localized to the ends of the system. These modes have their frequencies pinned to the middle of the band gap, even in the presence of disorder, as long as certain symmetries are preserved. In this case, topology provides a systematic way to create localized defect modes at a specific frequency. 

The second important class of topological modes are edge states of two-dimensional topological systems. These exhibit either unidirectional or spin-controlled propagation along the edge. The former appear in topological systems with broken time-reversal symmetry (known as quantum Hall topological phases~\cite{TKNN}), while in photonics the latter require a combination of time-reversal symmetry and some other internal symmetry protecting a spin-like degree of freedom (known as quantum spin-Hall topological phases~\cite{QSH}). In both cases the resulting topological edge states are of interest as a means of creating disorder-robust optical waveguides.

The generic approach to create a topological system is to is to start with a simple periodic medium exhibiting some degeneracy in its photonic band structure, and then break one of the symmetries protecting the degeneracy to open a band gap. Breaking the symmetry in the right way will create a topologically nontrivial band gap hosting protected edge modes.

\begin{figure}
    \includegraphics[width=\columnwidth]{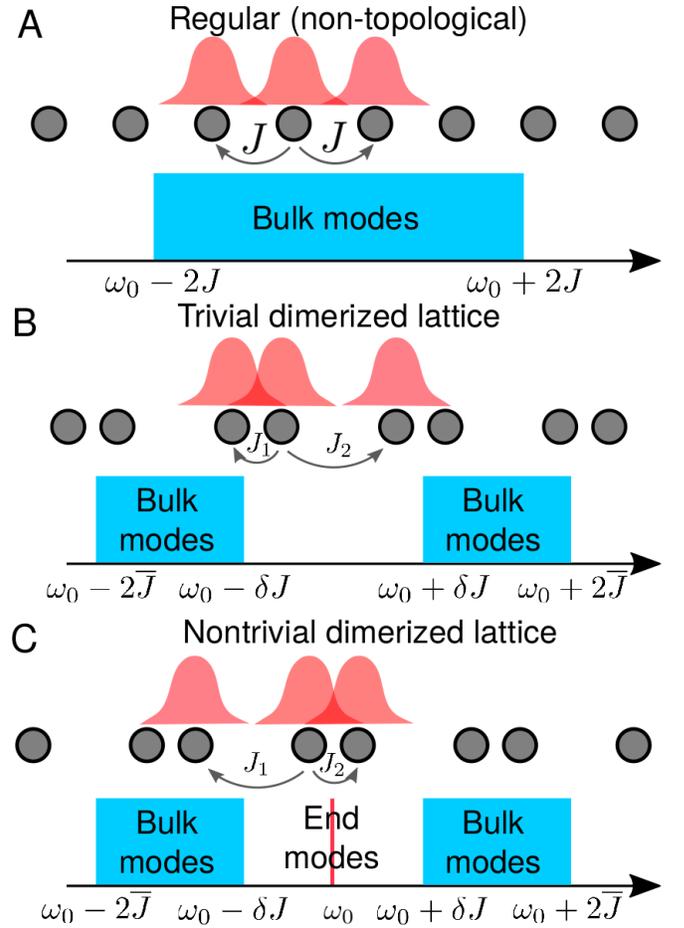}
    \caption{Schematic of general procedure for creating a topological medium. (A) A regular periodic lattice can be modelled as a collection of resonant elements (shaded circles) with individual resonant frequencies $\omega_0$ coupled together to form a band of delocalized bulk modes. \revision{The coupling strength $J$ is determined by the overlap between the modes of the individual lattice sites, sketched in red.} (B,C) Reducing the translation symmetry of the lattice \revision{by introducing staggered couplings $J_1 \ne J_2$} opens a mini-gap in the bulk spectrum \revision{of size $2\delta J$, where $\delta J = |J_1-J_2|$ and $\overline{J} = \frac{1}{2}(J_1+J_2)$}. (B) The trivial dimerized lattice $J_1 > J_2$ hosts two bands corresponding to symmetric and antisymmetric dimer modes, separated by a mini-gap. (C) In the nontrivial dimerized lattice $J_1 < J_2$ the dimers at the ends are broken, producing in mid-gap modes localized to the ends of the lattice.}
    \label{fig:topology}
\end{figure}

Fig.~\ref{fig:topology} shows a simple one-dimensional example of this idea, where one can reduce the translation symmetry of a periodic lattice by staggering its site positions to create a lattice of dimers. Reducing the translation symmetry opens a mini-gap the lattice's band structure, analogous to the Su-Schrieffer-Heeger tight binding Hamiltonian for electron transport in polymers~\cite{SSH,Asboth_book},
\revision{\be 
\hat{H}_{\mathrm{SSH}} = \sum_{n} \left(J_1 \hat{a}^{\dagger}_{2n-1} \hat{a}_{2n} + J_2 \hat{a}^{\dagger}_{2n} \hat{a}_{2n+1} \right) + \mathrm{c.c.},
\ee
where $\hat{a}_n$ is the annihilation operator for the $n$th site, and $J_{1,2}$ are coupling strengths. Note that while $\hat{H}_{\mathrm{SSH}}$ is written using second quantization notation, it is equally applicable to classical states of light, where the eigenvalues of $\hat{H}_{\mathrm{SSH}}$ give the frequency detunings of the collective array modes, as shown in Fig.~\ref{fig:topology}.}

\revision{There are two different ways to dimerize a finite lattice; we can have either $J_1 > J_2$ or $J_1 < J_2$, corresponding to two inequivalent topological phases distinguished by a quantized winding number. The trivial phase $J_1>J_2$ forms a set of dimers and does not exhibit any end modes. On the other hand, in the nontrivial phase $J_1 < J_2$ the dimers are broken at the ends of the system, resulting in a pair of localized end modes with frequencies lying in the middle of the band gap. The end modes are topologically protected in the sense that introducing disorder to the inter-site couplings does not shift their frequencies; they remain ``pinned'' to the middle of the band gap and localized to the ends as long as the disorder is sufficiently weak that the bulk band gap remains open.}

\revision{Thanks to its simplicity and relative ease of implementation, photonic analogues of the Su-Schrieffer-Heeger model have been realized in a variety of platforms~\cite{Ozawa2019}, including waveguide lattices~\cite{Malkova2009,Keil2013,BlancoRedondo2016}, plasmonic and dielectric nanoparticle arrays~\cite{Sinev2015,Kruk2017}, photonic crystals~\cite{Wang2016, Ota2018}, and microring resonators~\cite{Parto2018,Zhao2018}.}

To create two-dimensional topological phases there are two common approaches: using synthetic gauge fields or by perturbing honeycomb lattices. Synthetic gauge field refers to complex (direction-dependent) coupling between different sites of the photonic lattice, arising in systems with non-reciprocity or broken time-reversal symmetry, \revision{corresponding to tight binding coupling terms of the form $J e^{i \theta} \hat{a}_{n+1}^{\dagger} \hat{a}_n + J e^{-i \theta} \hat{a}_n^{\dagger} \hat{a}_{n+1}$, where $\theta$ is the coupling phase}. Complex coupling is formally equivalent to the effect of an electromagnetic vector potential on electron transport. Using a suitable position-dependent synthetic gauge field \revision{$\theta = \theta(x,y)$} allows one to create an effective magnetic field for light~\cite{Fang2012NP}, resulting in analogues of the quantum Hall topological phase~\cite{TKNN}.

The second approach to creating two-dimensional topological phases is based on the honeycomb lattice, which exhibits Dirac point degeneracies at the corners of its Brillouin zone. Weak symmetry-breaking perturbations are capable of lifting the degeneracy to open a topological band gap. One can break either time-reversal symmetry to create the quantum Hall phase~\cite{Haldane_model}, or other internal symmetries (e.g. related to sublattice or polarization degrees of freedom) to create quantum spin Hall phases~\cite{QSH}.

Other classes of topological models beyond the above gapped topological insulating phases are attracting increasing attention. For example, higher-order topological phases can give rise to modes localized to the corners of two-dimensional systems~\cite{quadrupole}. Three-dimensional Weyl topological phases exhibit protected degeneracies in their bulk band structure~\cite{weyl_review,weyl_photonic}. Non-Hermitian topological phases can emerge in systems with structured gain or loss~\cite{MartinezAlvarez2018,NH_1,NH_2,NH_3}. For further discussion on photonic topological phases we direct the reader to Refs.~\onlinecite{Ozawa2019,Khanikaev2017,Yuan2018,Ozawa2019NRP}.

It is important to stress that these topological photonic systems are only analogous to the topological tight binding models used to describe electronic condensed matter systems. Thus, while the electronic quantum Hall phase exhibits a Hall conductivity precisely quantized to 1 part in $10^9$, in photonics various effects such as material absorption, out-of-plane scattering, and imperfect symmetries mean that the topological protection is only approximate, so the edge modes are only protected against certain classes of perturbations. Thus, it is crucial to identify systems where topology provides protection against the most significant sources of disorder; most studies to date rely on deliberately-introduced defects to demonstrate topological protection. \revision{For example, the end states of the Su-Schrieffer-Heeger model are only protected against the ``off-diagonal'' disorder in the inter-site coupling coefficients, and are \emph{not} protected against the ``diagonal'' disorder in the individual sites' resonant frequencies, which leads to random variations to the end modes' resonant frequencies.}

\subsection{Ring resonators}
\label{sec:rings}

\begin{figure}

\includegraphics[width=\columnwidth]{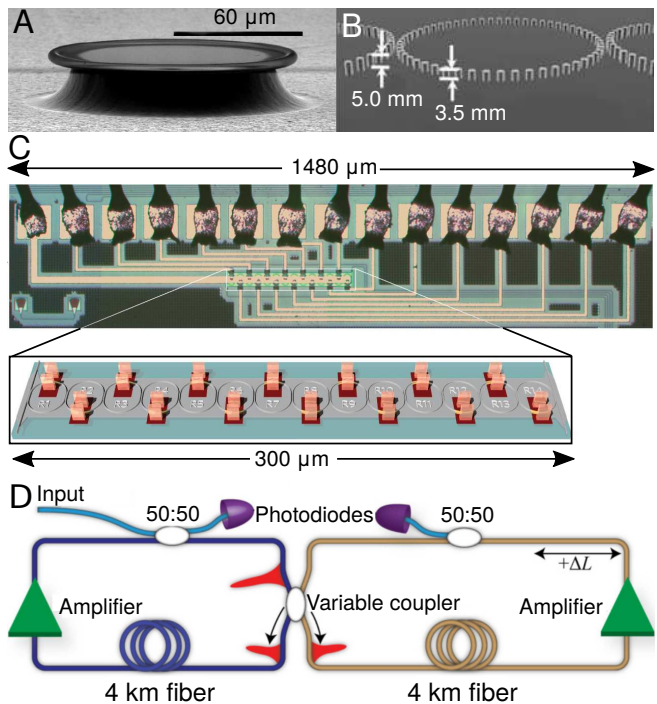}
    \caption{Examples of ring resonators in different platforms. (A) Silica microtoroid resonator, adapted from Ref.~\onlinecite{Armani2003}. (B) Coupled spoof plasmon ring resonators formed by subwavelength metal pillars, adapted from Ref.~\onlinecite{Gao2016}. (C) 14-ring CROW on silicon integrated with thermal tuners, adapted from Ref.~\onlinecite{Jayatilleka2019}. (D) Coupled fiber loops, \revision{where photodiodes are used to monitor the propagation dynamics by tracking the intensity within each loop}, adapted from Ref.~\onlinecite{Bisianov2019}. \revision{Experiments in panels (A,C,D) were conducted at telecom wavelengths ($\lambda \approx 1550$ nm), while (B) is a microwave frequency ($\omega \approx 11.3$ GHz) device.}}
    \label{fig:rings_intro}
\end{figure}

Ring resonator generally refers to any optical waveguide forming a closed loop, regardless of its size or shape~\cite{Bogaerts2012}. Fig.~\ref{fig:rings_intro} presents some examples of ring resonators, including micrometer-scale integrated optical resonators, millimeter-scale spoof plasmon resonators, and kilometer-scale fiber loops. Microwaves and fiber loops provide a convenient setting for studying novel design approaches as they are easier to fabricate, and in some cases can be assembled using off-the-shelf optical components, while integrated photonic circuits are of more interest for potential applications due to their compactness and scalability.

Resonances occur whenever the propagation phase accumulated over a round trip forms a multiple of $2\pi$. Key characteristics of ring resonators are their resonance width, free spectral range (FSR; the spacing between neighbouring resonances), quality factor (resonance frequency divided by width), and finesse (free spectral range divided by resonance width). The resonance width is dictated both by the intrinsic losses due to waveguide bending, scattering losses due surface roughness, absorption, and extrinsic losses introduced by coupling the resonator to external waveguides. \revision{We emphasize that in passive systems topological designs generally do not provide protection against these sources of loss.}

The free spectral range is inversely proportional to the round trip path length. In systems with small FSR such as long fiber loops one typically studies the propagation dynamics in the time domain. For on-chip signal processing applications it is desirable to have a large free spectral range, exceeding the signal bandwidth, demanding a high refractive index contrast to minimise waveguide bending losses. The resulting strong light confinement in turn makes the ring's resonances highly sensitive to local perturbations to the refractive index, which can be both a strength and a weakness. For example, the sensitivity to perturbations allows ring resonators to be employed as highly compact and efficient sensors and optical switches~\cite{Armani2006,Sun2011,Xu2014,Jayatilleka2018}. On the other hand, for spectral filtering applications active tuning is typically required to keep the resonance fixed at the desired frequency~\cite{Nawrocka2006}.

A high quality factor is desirable for nonlinear optics applications. \revision{For example, an intensity-dependent refractive index enables bistability as the input frequency is tuned, which is useful for all-optical switching~\cite{Almeida2004}. An intensity-dependent refractive index can arise not only due to the optical Kerr effect, but also from thermal nonlinearities and free carrier dispersion~\cite{Bogaerts2012}.} The differing characteristic time scales of these nonlinear effects can lead to complex pulsating dynamics~\cite{Priem2005}. Ultra-fast Kerr nonlinearities are also employed for frequency mixing applications, where the relatively uniformly spacing of the ring resonances are ideal for frequency comb generation~\cite{Kippenberg2018,Vasco2019,Drake2019,Chang2020,Riemensberger2020}.

The strong dispersion close to resonance allows ring resonators to be used to delay and store optical signals. For single rings there is a trade-off between the delay time and the operating bandwidth (their product is a fixed constant). Larger delays for a fixed bandwidth can be achieved using arrays of coupled rings, known as coupled resonator optical waveguides (CROWs)~\cite{Yariv99,Canciamilla10,Morichetti12,Takesue13}. In integrated photonics, however, disorder \revision{in the form of nanometer-scale variations to the rings' height and thickness} leads to significant misalignment of the individual rings' resonance frequencies, severely degrading the CROW performance~\cite{Cooper10}. Therefore one requires either clever designs that are robust against these fabrication variations~\cite{Ouyang2019} or active tuning to compensate for the disorder~\cite{Sokolov2017,Jayatilleka2019}.

The thermo-optic effect is the most common way to tune integrated photonic ring resonators, using micro-heaters placed on top of the individual rings. While thermal tuners offer a large tuning range, they are slow (operating on the $\mu$s scale), have poor energy efficiency~\cite{Zhou2015}, require careful design to minimise cross-talk between different tuners~\cite{Jacques2019}, and inevitably introduce sensitivity to environmental temperature fluctuations~\cite{Guha2010}. Electro-optic tuning can operate much faster (sub ns) and with greater energy efficiency, however the tuning range is smaller and additional absorption losses are introduced~\cite{Bogaerts2012,Timurdogan2014}. 

There is growing interest in methods to improve the reliability of ring resonator-based devices, such as creating temperature-insensitive resonators by combining materials with opposite thermo-optic coefficients~\cite{Guha2013}, and introducing tunable backscattering to cancel out backscattering caused by fabrication imperfections via destructive interference~\cite{Li2017}. \revision{Topological designs are a promising alternative exhibiting \emph{passive} robustness against disorder. For example, the topological edge modes of two-dimensional lattices are robust against the ``diagonal'' disorder formed by misalignment in the rings' resonance frequencies, provided the disorder strength does not exceed the size of the topological band gap.}

For further in-depth discussion of the physics and applications of photonic ring resonators we recommend Refs.~\onlinecite{Bogaerts2012,Morichetti12,Li2019}.

\section{Topological coupled resonator lattices}
\label{sec:lattices}

Arrays of coupled ring resonators provide a flexible platform for implementing topological lattice models. In weakly coupled arrays light propagation is governed by effective tight binding Hamiltonians which describe the evanescent coupling of light between neighbouring resonators. The magnitude of the coupling coefficients can be controlled simply by varying the separation between the resonators. Furthermore, coupling resonant rings via anti-resonant links allows one to tune the phase of the coupling. One can effectively break time-reversal symmetry by considering modes with a fixed ``spin'' (clockwise or anti-clockwise circulation direction) and neglecting backscattering within the rings. On the other hand, taking this backscattering into account introduces an in-plane effective magnetic field~\cite{Hafezi2011}. Together, these ingredients enable the realization of a wide variety of topological tight binding models in one and two dimensions.

\begin{figure}

\includegraphics[width=\columnwidth]{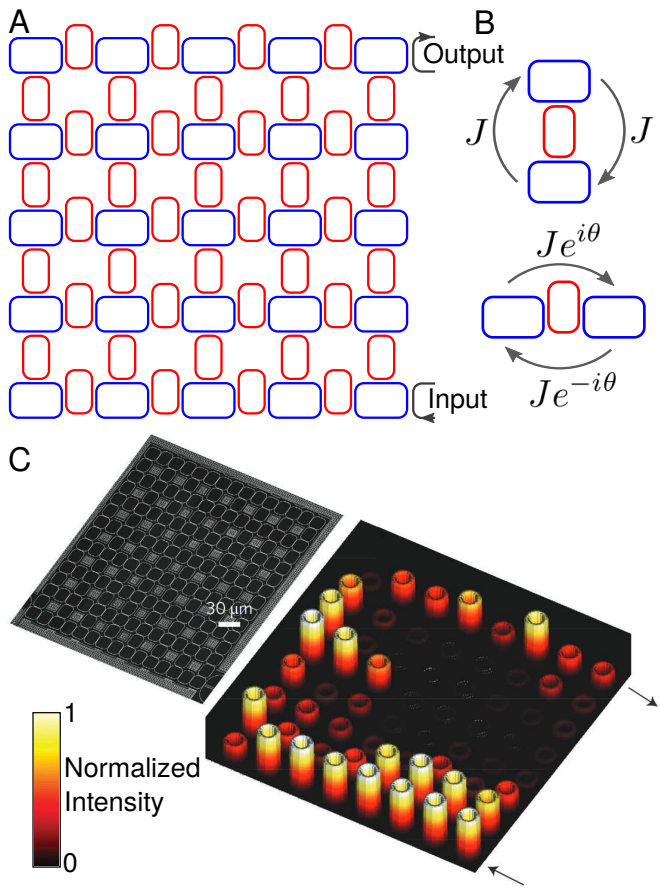}

\caption{Synthetic magnetic field in a ring resonator lattice. (A) Schematic of lattice, consisting of resonant ``site'' rings (blue) coupled via anti-resonant ``link'' rings (red). Light is injected at the input into anticlockwise-circulating site modes, effectively breaking time-reversal symmetry. (B) Schematic of coupling terms in the tight binding Hamiltonian. Coupling in the vertical direction is symmetric, while coupling in the horizontal direction is accompanied by a hopping phase $e^{\pm i \theta}$ due to the asymmetry of the link ring. An inhomogeneous hopping phase $\theta = \alpha y$ induces a synthetic magnetic flux $\alpha$. (C) Image of the first photonic topological resonator lattice and propagation of its edge states, adapted from Ref.~\onlinecite{Hafezi2013}.}

\label{fig:QHE}

\end{figure}

Studies of topological ring resonator lattices began with the seminal theoretical work of Hafezi et al.~\cite{Hafezi2011}, which showed that one can effectively break time-reversal symmetry by exciting modes with a particular circulation direction and then use asymmetric link rings to implement an analogue of the quantum Hall lattice model. The asymmetric link rings result in a phase difference between the two coupling directions, equivalent to a vector potential in the tight binding Hamiltonian. By making this hopping phase inhomogeneous (e.g. proportional to the $y$ coordinate) one can create a vector potential formally equivalent to an out-of-plane effective magnetic field, which implements a lattice model of the quantum Hall effect, as shown in Fig.~\ref{fig:QHE}(A,B) \revision{and described by the tight binding Hamiltonian~\cite{Hafezi2011}
\begin{align} 
\hat{H}_{\mathrm{QH}} = J \sum_{x,y} \left( \hat{a}_{x+1,y}^{\dagger} \hat{a}_{x,y} e^{-i \alpha y} + \hat{a}_{x,y}^{\dagger} \hat{a}_{x+1,y} e^{i \alpha y} \right. \nonumber \\ \left. + \hat{a}_{x,y+1}^{\dagger} \hat{a}_{x,y} + \hat{a}_{x,y}^{\dagger} \hat{a}_{x,y+1} \right),
\end{align}
where $\alpha$ is the effective magnetic flux threading each plaquette of the square lattice.}

The topology of the quantum Hall lattice is characterized by the quantized Chern number, which determines the number of chiral backscattering-protected states at the edge of the lattice. \revision{Specifically, for a given band gap one sums the Chern numbers of all the bulk bands below the gap to obtain the gap Chern number~\cite{Ozawa2019}. The number of chiral edge states at an interface between two media is then given by the difference between the gap Chern numbers of the two media.} Note that as this \revision{resonator lattice} obeys time-reversal symmetry the opposite spin hosts counter-propagating edge states (forming a quantum spin Hall phase), and hence the protection against backscattering only holds as long as the two spins remain decoupled. 

Following this proposal the first experiment was reported in 2013~\cite{Hafezi2013}. \revision{The experiment was performed in the telecom band ($\lambda \approx 1550$ nm) and used a lattice with FSR $\approx 10^3$ GHz and inter-site coupling strength $J \approx 16$ GHz, deep in the tight binding regime. The diagonal disorder in the rings' resonant frequencies estimated to be $\approx 0.8J$, smaller than the size of the topological band gap. Other forms of disorder were found to be negligible for the system parameters considered: the strength of the inter-site coupling disorder and intra-site coupling between the two spins were both estimated to be $0.04J$.} 

In contrast to the ideal lossless case, in practice the individual rings exhibited intrinsic losses \revision{$\kappa_{\mathrm{in}} \approx 1$ GHz} due to roughness of the waveguide walls and absorption. This sets an upper limit on the propagation length of the topological edge states despite their protection against backscattering induced by disorder in the rings' resonant frequencies \revision{and inter-ring couplings}. These losses were harnessed in the experiment to directly image the propagation of the topological edge states by measuring the light scattered out of the device plane.

Fig.~\ref{fig:QHE}(C) shows the device and an image of the topological edge states, which reliably travel from the input port to an output port. In contrast the bulk states exhibited Anderson localization due to the strong intrinsic disorder present in the system. Subsequent experiments measured the delay times through several devices, showing indeed that the topological edge states preserve ballistic light transport with low device-to-device fluctuations in the photon delay times, whereas regular CROWs exhibit strong fluctuations due to the disorder-induced scattering~\cite{Mittal14}. The quantum Hall lattice model was also implemented using silicon nitride ring resonators, however the propagation distance of the topological edge states was limited by stronger intrinsic losses \revision{$\kappa_{\mathrm{in}} \approx J \approx 60$ GHz}~\cite{Yin2016}.

Other topological tight binding models have also been studied using ring resonator lattices. A higher order topological phase exhibiting protected corner states was demonstrated in 2019~\cite{Mittal2019b}. Next-nearest neighbour coupling was used to implement an analogue of the Haldane model~\cite{Haldane_model}, which exhibits quantum Hall edge states even in the absence of a net effective magnetic flux~\cite{Leykam2018,Mittal2019}. Zhu et al. have proposed a honeycomb lattice design hosting topological edge states which co-exist with a nearly flat bulk band~\cite{Zhu2018}. It is also possible to shrink these two-dimensional lattices down to quasi-one-dimensional delay lines, which maintain some resistance against disorder~\cite{Han19}.

In 2018 the ring resonator platform was used to implement one- and two-dimensional topological laser models by embedding a quantum well gain medium into the resonators~\cite{Parto2018,Zhao2018,Bandres2018,Harari2018}. One-dimensional experiments were carried out using the Su-Schrieffer-Heeger lattice, created by staggering the separation between neighbouring rings. Pumping one of two sublattices comprising the array induced lasing of its mid-gap topological edge states~\cite{Parto2018,Zhao2018}. Two-dimensional lasing experiments utilised the quantum Hall lattice, where a pump localized to the lattice edges induced lasing in its chiral edge states~\cite{Bandres2018,Harari2018}. In both cases, the potential advantage of the topological approach is the ability to induce lasing in collective array modes that are \revision{localized by the topological band gap}. For further information on topological lasing, we recommend the recent reviews Refs.~\onlinecite{active_review,nonlinear_review}. 

One advantage of the silicon photonics platform is the ability to implement actively-tunable devices, for example by incorporating thermo-optic phase shifters to tune the resonant frequencies of the individual rings. Mittal et al.~\cite{Mittal2016} employed tunable phase shifters at the edge of the quantum Hall lattice to directly measure the topological winding number of the edge states. Similar tuning of the effective magnetic flux in ring-shaped lattices enables the observation of the Hofstadter butterfly via the lattice's scattering resonances~\cite{Ao2018,Zimmerling2020}. There are recent proposals to implement phase shifters throughout the entire lattice in order to tune its topological properties, thereby enabling one to switch the topological edge states on or off or re-route them between different output ports~\cite{Leykam2018,Kudyshev2019thermal,Kudyshev2019oxides}.  Zhao et al. demonstrated controllable re-routing of topological states in the quantum Hall resonator lattice using structured bulk gain~\cite{Zhao2019}.

\begin{figure}
    \centering
    \includegraphics[width=\columnwidth]{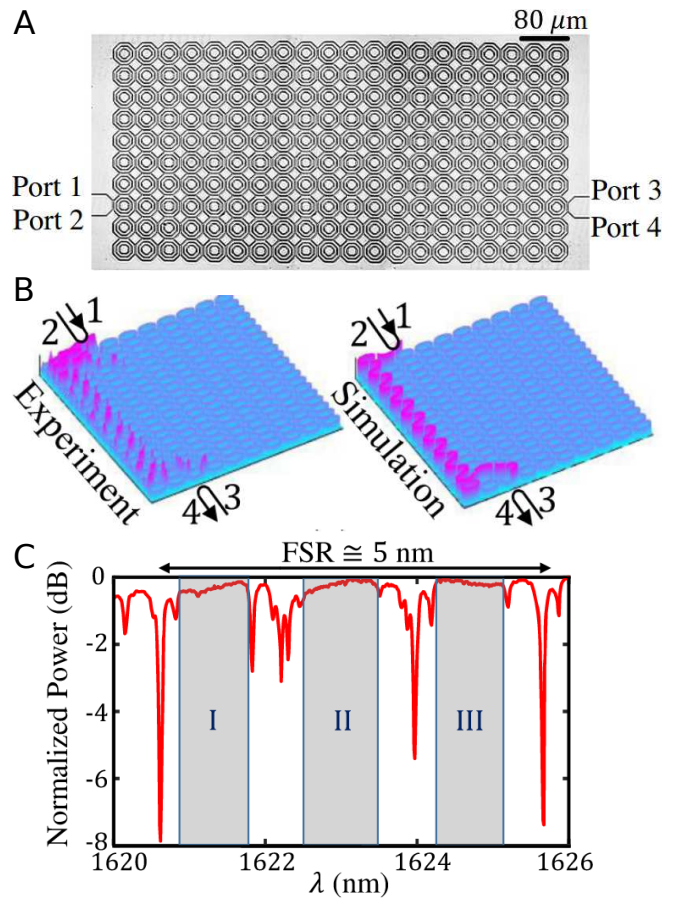}
    \caption{Anomalous Floquet topological resonator lattice, adapted from Ref.~\onlinecite{Afzal2019}. (A) Microscope image of the silicon resonator lattice. (B) Image of scattered light when the chiral edge state is excited (left), compared with the simulated edge state intensity distribution (right). (C) Measured transmission spectrum of the lattice, \revision{with high transmission in all three band gaps (shaded grey regions I, II, III) mediated by the anomalous Floquet topological edge states.}}
    \label{fig:AFI}
\end{figure}

The above studies of topological resonator lattices focused on the weak coupling limit described by tight binding models. However, topological phases can also arise in the strongly coupled lattices, without requiring external modulation or anti-resonant link rings~\cite{Chong2013,Pasek2014,Shi2017,Afzal2018}. These ``anomalous Floquet'' phases are not predicted by the tight binding approximation and only emerge when considering the full transfer matrix description of the light coupling between neighbouring rings. A ring resonator lattice implementing an anomalous Floquet topological phase was first demonstrated in 2016 using spoof plasmons at microwave frequencies~\cite{Gao2016}. Similar anomalous edge states can emerge in weakly coupled arrays with gain and loss, where the transfer matrix description is essential to account for the growth or decay of the optical field as it circulates through each ring~\cite{Ao2020}. 

The anomalous Floquet topological phase was scaled up to telecom wavelengths by Afzal et al. using the silicon photonic resonator lattice shown in Fig.~\ref{fig:AFI}~\cite{Afzal2019}. The main distinguishing feature compared to previous observations of quantum Hall edge states is the presence of edge states in all of the array's band gaps. One advantage of the strongly-coupled anomalous Floquet phases is that their bulk bands and edge states can have bandwidths comparable to the rings' free spectral range, in contrast to tight binding lattices which are typically restricted to small bandwidths. \revision{However, the stronger coupling implies a reduction in the rings' quality factors and hence suppression of nonlinear effects. Thus, whether it is better to employ anomalous Floquet-type lattices or topological tight binding lattices will depend on the particular application.}

\section{Topology of dynamically-modulated resonators}
\label{sec:modulation}

Resonators undergoing the dynamic modulation of the refractive index provide a flexible way to construct effective lattice models described by time-dependent tight binding Hamiltonians. Such systems break time-reversal symmetry and provide an important platform for implementing various topological phenomena. This specific subject was started with the pioneering paper by Fang et al.~\cite{Fang2012PRL}, who showed that in photonic systems where the refractive index is harmonically modulated, the modulation phase actually gives rise to an effective gauge potential for photons. Based on this idea, a photonic Aharonov-Bohm interferometer was proposed as a design for an optical isolator. 

\begin{figure}
\centering
\includegraphics[width=\columnwidth]{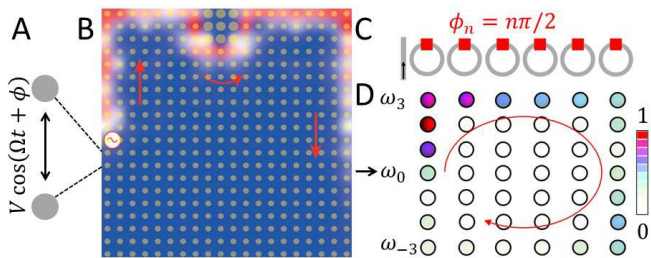}
\caption{Dynamically-modulated resonators generating effective magnetic fluxes. (A) Two resonators are connected through the dynamic modulation. (B) Topologically-protected one-way edge mode propagates around the defect in a two-dimensional resonator lattice, where the spatial distribution of modulation phases gives an effective magnetic field. (C) An array of rings undergoing dynamical modulation, which creates a synthetic space (D) with one dimension being the frequency dimension, where a topological one-way edge mode can be excited. (A,B) are adapted from Ref.~\onlinecite{Fang2012NP}. (C,D) are adapted from Ref.~\onlinecite{Yuan2016OL}. \label{fig:DMR} }
\end{figure}

A subsequent work by Fang et al. proposed a scheme for generating an effective magnetic field for photons~\cite{Fang2012NP}, based on spatially inhomogeneous modulation phases. The effective magnetic field for photons breaks time-reversal symmetry and can be used to induce nontrivial quantum Hall phases in two-dimensional resonator lattices. Topologically-protected one-way edge modes can be excited in this dynamically-modulated resonator lattice which are robust against defects, as shown in Fig.~\ref{fig:DMR}. In contrast to the lattices discussed in the previous section, these topological modes are also robust against spin-flipping disorders \revision{because time-reversal symmetry is broken. However, they remain susceptible to intrinsic losses such as absorption.}

Experiments based on these ideas were implemented in a variety of platforms. The first proof-of-principle demonstration of a photonic Aharonov-Bohm interferometer used an electrical network at radio frequencies~\cite{Fang2013PRB}. In 2014, the photonic Aharonov-Bohm effect has been demonstrated at visible wavelengths by utilizing an effective gauge potential induced by photon-phonon interactions~\cite{Li2014NC}. Later in the same year, the presence of an effective gauge field has been constructed using the on-chip silicon photonics technology, where the refractive index of the silicon coupled waveguides was modulated by an applied voltage~\cite{Tzuang2014NP}. These works are important proofs-of-concepts for photonic gauge potentials induced by dynamic modulation.

These proposals for creating effective gauge potentials in dynamically-modulated resonators have triggered many follow-up studies on the manipulation of light via modulation phases. For example, spatially inhomogeneous distribution of modulation phases in two-dimensional resonator lattices can be used to control the flow of light~\cite{Fang2013PRL,Fang2013OE,Lin2014PRX,Minkov16}. A spatially homogeneous but time-dependent distribution of modulation phases in three dimensions has also been considered, which results in propagation analogous to the dynamics of electrons in the presence of a time-dependent electric field~\cite{Yuan2015PRL}. By temporally modulating the effective electric field, one can time-reverse the propagation of one-way quantum Hall edge states~\cite{Yuan2016PRB}.

In the above studies the modulations under considerations were treated weakly, so that the dynamics satisfy the rotating-wave approximation. Topological phase transitions have also been studied in the ultrastrong coupling regime, where the rotating wave approximation fails. In the ultrastrong coupling regime the topological edge modes have been shown to exhibit larger bandwidth and less susceptibility to losses~\cite{Yuan2015PRA}. On the other hand, experimental efforts are still ongoing to achieve ultrastrong coupling using ring resonators. As an experimental proof of concept, light guiding by an effective gauge potential~\cite{Lin2014PRX} has been demonstrated in tilted waveguide arrays~\cite{Lumer2019NP}. Moreover, lithium niobate microring resonators have been coupled and modulated by external microwave excitation, which leads to an effective photonic molecule~\cite{Zhang2019NP}. Various platforms have therefore been shown as potential candidates for exploring resonators under strong dynamic modulation.

Besides studying topological physics in real space, dynamically modulated resonators also provide a unique platform to explore higher-dimensional topological physics in lower dimensional physical systems, by incorporating synthetic dimensions in photonics~\cite{Yuan2018,Ozawa2019NRP}. Inspired by earlier works of synthetic dimensions in lattice systems~\cite{Tsomokos2010PRA,Boada2012PRL,Jukic2013PRA}, resonators supporting multiple degenerate modes with different orbital angular momentum (OAM) have been used to simulate the topological physics, where the synthetic dimension is constructed by coupling modes with different OAM using a pair of spatial light modulators~\cite{Luo2015NC,Sun2017PRA,Zhou2017PRL,Luo2018PRA}. 

On the other hand, dynamically-modulated ring resonators with the modulation frequency close to the resonators' free spectral range naturally gives rise to a synthetic dimension along the frequency axis of light~\cite{Yuan2016OL,Ozawa2016}. Using this idea, two-dimensional topologically-protected one-way edge states have been proposed using one-dimensional resonator lattices [see Fig.~\ref{fig:DMR}(C)]. Such edge modes convert the frequency of light unidirectionally towards higher (or lower) frequency components as shown in  Fig.~\ref{fig:DMR}(D), which could form the basis for a topological frequency converter~\cite{Yuan2016OL}. The four-dimensional quantum Hall effect can also be studied using this approach, by combining a three-dimensional resonator lattice with a fourth synthetic frequency dimension~\cite{Ozawa2016}.

Synthetic dimensions in dynamically-modulated resonators also provide a platform for exploring novel topological phases that are difficult to implement using pure spatial lattices. For example, using synthetic dimensions it is possible to implement three-dimensional Weyl~\cite{Lin2016NC,Zhang2017PRA} and topological insulating phases~\cite{Lin2018SA} using two-dimensional arrays of rings. In two-layer two-dimensional ring lattices, higher-order topological phases exhibiting corner states have also been designed~\cite{Dutt2019arxiv}. Based on the scheme of creating topological system in a one-dimensional array of ring resonators, a mode-locked topological insulator laser in synthetic dimensions has been suggested, which triggers potential applications for developing active photonic devices~\cite{Yang2020}. 

One significant advantage of synthetic dimensions implemented using dynamically modulated resonators is the ability to flexibly control the connectivity of the couplings in the synthetic space, which is difficult to achieve in real space lattices~\cite{Schwartz2013OE,Bell2017,Titchener2020}. For example, one can introduce long-range couplings along the synthetic frequency dimension by using modulation frequencies that are multiples of the FSR, enabling emulation of the two-dimensional Haldane model using three rings~\cite{Yuan2018PRB}. Moreover, in a single resonator, one can combine two internal degrees of freedom of light such as frequency and OAM to construct a two-dimensional synthetic lattice~\cite{Yuan2019PRL}. In such synthetic lattices, the effective magnetic field can be naturally introduced through the additional coupling waveguides, thereby creating topologically-protected one way edge states. This may enable the robust manipulation of entanglement between multiple degrees of freedom of light.

Besides topological physics, many other interesting analogies with quantum and condensed matter physics can be demonstrated using dynamically-modulated resonators, including Bloch oscillations~\cite{Longhi2005OL,Yuan2016Optica}, parity-time symmetric systems~\cite{Longhi2016OL,Yuan2018APLP}, and flatband lattices~\cite{LonghiABcage,Yu2020arxiv}. Furthermore, the creation of local nonlinearity in the synthetic frequency dimension are under study, which could significantly broadens the range of Hamiltonians involving local interactions that can be considered in the photonic synthetic space in dynamically-modulated resonators~\cite{Yuan2019arxiv}.

\begin{figure}
\centering
\includegraphics[width=\columnwidth]{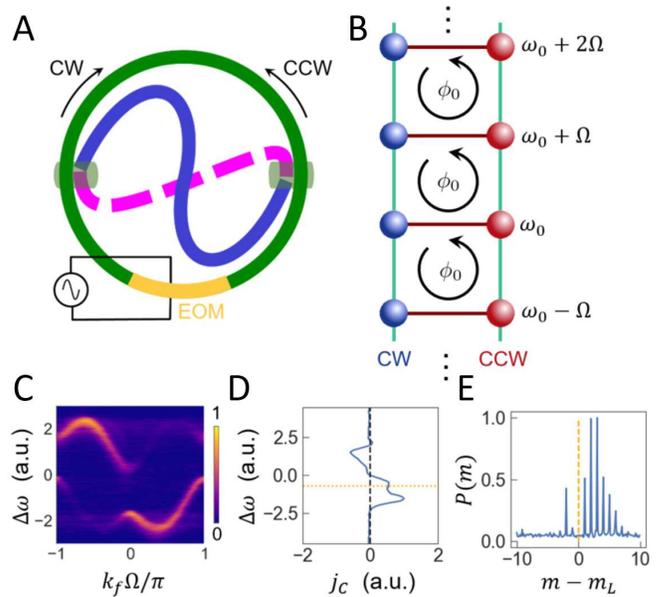}
\caption{Topology in synthetic space, adapted from Ref.~\onlinecite{Dutt2020}. (A) Schematic of a single ring formed by a fiber loop under the dynamic modulation. The CW and CCW modes \revision{form a spin degree of freedom and} are coupled via connecting waveguides of differing lengths. (B) The corresponding lattice gives a synthetic Hall ladder \revision{threaded by an effective magnetic flux $\phi_0$} with two independent synthetic dimensions. (C) Measuring the transmission of an input CW excitation as a function of the detuning $\Delta \omega$  reveals two bands with opposite chirality. (D) The chiral current $j_C$, \revision{a measure of the spin-sensitive direction of frequency conversion, is opposite for the two bands}. (E) Steady state normalized photon number $P(m)$ of the CW mode at the $m$-th resonance, \revision{where $m_L$ is the resonant mode closet to the input laser, indicating preferential conversion to higher frequencies.}\label{fig:DMR2} }
\end{figure}

We conclude this Section by discussing some recent experimental demonstrations of synthetic dimensions in photonics. The first photonic topological insulator in a synthetic dimension was demonstrated using an array of multimode waveguides, where modulation of the refractive index along the waveguide axis played the role of the dynamic modulation~\cite{Lustig2019Nature}. The dynamically-modulated resonator has also been implemented in the fiber-based ring experiments incorporating commercial electro-optic modulators~\cite{Chen2018,Chalabi2019}, where band structures associated with one-dimensional synthetic lattices along the frequency axis of light have been measured~\cite{Dutt2019}. Based on this experimental setup, one can use the clockwise/counter-clockwise modes of a single ring as another degree of freedom to construct a synthetic Hall ladder with two independent physical synthetic dimensions, as shown in Fig.~\ref{fig:DMR2}~\cite{Dutt2020}. An effective magnetic flux was generated in the experiment and signatures of topological chiral one-way edge modes were observed. Recently, integrated lithium niobate resonators under dynamic modulation provide another potential experimental platform to construct synthetic dimensions and explore topological photonics in a synthetic space, which is potentially significant for on-chip device applications~\cite{Reimer2019arxiv}.

\section{Future directions}
\label{sec:future}

Having reviewed some of the seminal works on implementing topological effects using ring resonators, we now discuss some promising directions for future research, including fundamental studies of topological phenomena and practical problems which must be solved to make topological ring resonators viable for device applications. 

Most studies of topological ring resonators to date have focused on Hermitian topological lattice models, in which strictly speaking the topological protection only holds in the absence of any gain or loss. The study of non-Hermitian topological phases induced by appropriately-structured gain or loss is a topic attracting enormous interest nowadays~\cite{MartinezAlvarez2018,NH_1,NH_2,NH_3}. Non-Hermitian coupling, which can induce novel non-Hermitian topological phases, can be implemented in coupled resonator lattices either by introducing asymmetric backscattering to the site rings~\cite{Malzard2015,Malzard2018} or adding gain or loss into the links to induce a hopping direction-dependent amplification or attenuation~\cite{Longhi2015,Leykam2017}. The latter has been implemented using variable-gain amplifiers in a microwave network~\cite{Hu2017} and coupled fiber loops~\cite{funneling}. Experiments with microring resonators remain limited to topological laser experiments based on adding gain to existing Hermitian topological phases~\cite{Parto2018,Zhao2018,Bandres2018,Harari2018}, making this an interesting direction for further studies. Can we use ideas from non-Hermitian topological phases to exploit or minimize the scattering losses present in integrated photonic ring resonators? 

Ring resonators also provide an ideal platform for studying nonlinear topological systems~\cite{nonlinear_review}, due to the enhancement of nonlinear effects provided by high quality factor microresonators. Recent experiments have harnessed nonlinearities to generate frequency combs from single resonators~\cite{Kippenberg2018,Vasco2019,Drake2019,Chang2020,Riemensberger2020}. It will be interesting consider topological band structure effects in this context. The high flexibility in controlling the inter-ring coupling in resonator lattices also provides an opportunity to implement exotic forms of nonlinearity, such as models with nonlinear coupling~\cite{Menotti19}. Systems with nonlinear coupling can exhibit nonlinearity-induced topological transitions~\cite{Hadad2016}, which were so far limited to electronic circuit experiments~\cite{Hadad2018}. Fiber loops are another promising platform for exploring nonlinear effects due to their long accessible propagation lengths and ability to compensate for losses using fiber amplifiers~\cite{Bisianov2019}.

Recently, several studies have proposed the use of topological modes supported by domain walls of topological photonic crystals as a means of constructing novel classes of ring resonators~\cite{Yang2018,Smirnova2019,Jalali2020,Barik2020}. Light confinement is typically weaker than the standard approach based on integrated photonic ridge waveguides, meaning larger resonator sizes are required. A potential benefit of topological photonic crystal-based ring resonators is their ability to support sharp corners without bending losses. However, it remains to be seen whether they will be competitive with existing ring resonators. For example, topological photonic crystals have been shown to exhibit large losses ($>100$ dB/cm) compared to conventional photonic crystal waveguides (5 dB/cm), due to out-of-plane scattering losses~\cite{Sauer2020}.

Research on topological photonics has so far largely focused on the fundamental science and demonstration of novel topological effects. There remains a large gap between these studies and potential applications which must be bridged. While new kinds of topological phenomena such as higher order topological phases~\cite{quadrupole,Mittal2019b} continue to attract fundamental interest, the need for more challenging ingredients such as high dimensional lattices or protecting symmetries makes any useful applications a far off prospect at this stage. Moving forward, we will require better optimization of existing topological designs to make them more competitive with standard components, moving from a paradigm of demonstrating topological robustness by deliberately introducing defects (as is the case in most experiments utilising photonic crystals, waveguide arrays, and metamaterials), to one where there is topological protection against the actual imperfections which limit the performance of real devices. Topological ring resonator lattices using silicon photonics are noteworthy as they are perhaps the only platform to date in which the topology imbues protection against the dominant form of intrinsic disorder, i.e. the misalignment in the rings' resonant frequencies.

Many applications of ring resonators employ small systems consisting of up to a few coupled rings. Generally for topological protection to hold, we need to have a bulk, requiring a large system size. Intuitively, it is the presence of a bulk which allows signals to route around imperfections on the edge. So another important direction is to determine how to use topological ideas to improve the performance of small systems of a few coupled resonators. This is a direction where concepts such as synthetic dimensions will likely play a key role.

Finally, one of the most exciting potential near-term applications of topological resonator lattices is as reliable delay lines or light sources in large scale quantum photonic circuits. For example, topological edge states may be useful as disorder-robust delay lines for entangled states of light~\cite{Mittal16Entangled,Rechtsman2016,Han2020}. In 2018, Mittal et al. demonstrated experimentally the generation of correlated photon pairs via spontaneous four wave mixing in a topological edge mode~\cite{Mittal2018}. They observed better reproducibility of the photon spectral statistics over several devices compared to regular CROWs, which is promising for the scaling up and mass production of quantum photonic circuits. Very recently this idea was generalized to dual pump spontaneous four wave mixing, which allows one to tune the resulting two photon correlations by changing the pump frequencies~\cite{Orre2020}.

\section{Conclusion}
\label{sec:conclusion}

We have presented an overview of how ring resonators provide a highly flexible platform for studying topological band structure effects in photonics. Ring resonators have not only enabled the implementation of seminal topological lattice models from condensed matter physics, but have also been used for some of the first observations of novel topological phases in any platform, such as higher order topological corner states. There is now strong theoretical and experimental evidence that topological ideas may useful for designing superior delay lines or frequency-converters in integrated photonic circuits. As the basic concepts are now well-established, future research will need to shift focus towards optimization of existing topological ring resonator systems to improve their performance and make their figures of merit more competitive with conventional ring resonator-based components. 

\section*{Acknowledgements}

We thank M. Hafezi for useful discussions. This research was supported by the Institute for Basic Science in Korea (IBS-R024-Y1, IBS-R024-D1), the National Natural Science Foundation of China (11974245), and the Natural Science Foundation of Shanghai (19ZR1475700).

\section*{References}

\bibliography{RingResonators}

\end{document}